\documentclass[a4paper,twocolumn,11pt,accepted=2021-07-22]{quantumarticle}
\pdfoutput=1
\usepackage[utf8]{inputenc}
\usepackage[english]{babel}
\usepackage[T1]{fontenc}
\usepackage{amsmath}

\usepackage{tikz}
\usepackage{lipsum}

\usepackage{custom}

\usepackage{dsfont}
\usepackage{pifont}
\usepackage{diagbox}
\usepackage{hyperref}
\usepackage{xcolor}
\definecolor{Bblue}{rgb}{0.19, 0.55, 0.91}

\newcommand{\ketbra}[2]{| {#1} \vphantom{#2} \rangle\langle {#2} \vphantom{#1} |} 
\newcommand{\proj}[1]{\ketbra{#1}{#1}}
\newcommand{\da}{\downarrow}
\newcommand{\ua}{\uparrow}
\DeclareMathOperator*{\ox}{\otimes}

\begin{document}

\title{Generalized probability rules from a timeless formulation of Wigner's friend scenarios}

\author{Veronika Baumann}
\thanks{The first three authors contributed equally.}
\affiliation{Institute for Quantum Optics and Quantum Information (IQOQI-Vienna) of the Austrian Academy of Sciences, Boltzmanngasse 3, A-1090, Vienna, Austria}
\affiliation{Faculty of Physics, University of Vienna, Boltzmanngasse 5, A-1090 Vienna, Austria}
\affiliation{Faculty of Informatics, Università della Svizzera italiana, Via G. Buffi 13, CH-6900 Lugano, Switzerland}

\author{Flavio Del Santo}
\affiliation{Institute for Quantum Optics and Quantum Information (IQOQI-Vienna) of the Austrian Academy of Sciences, Boltzmanngasse 3, A-1090, Vienna, Austria}
\affiliation{Faculty of Physics, University of Vienna, Boltzmanngasse 5, A-1090 Vienna, Austria}

\author{Alexander R. H. Smith}
\orcid{0000-0002-4618-4832}
\affiliation{Department of Physics and Astronomy, Dartmouth College, Hanover, New Hampshire 03755, USA}

\author{Flaminia Giacomini}
\affiliation{Institute for Quantum Optics and Quantum Information (IQOQI-Vienna) of the Austrian Academy of Sciences, Boltzmanngasse 3, A-1090, Vienna, Austria}
\affiliation{Faculty of Physics, University of Vienna, Boltzmanngasse 5, A-1090 Vienna, Austria}
\affiliation{Perimeter Institute for Theoretical Physics, 31 Caroline St. N, Waterloo, Ontario, N2L 2Y5, Canada}

\author{Esteban Castro-Ruiz}
\affiliation{Institute for Quantum Optics and Quantum Information (IQOQI-Vienna) of the Austrian Academy of Sciences, Boltzmanngasse 3, A-1090, Vienna, Austria}
\affiliation{Faculty of Physics, University of Vienna, Boltzmanngasse 5, A-1090 Vienna, Austria}
\affiliation{QuIC, Ecole polytechnique de Bruxelles, C.P. 165, Universit\'e libre de Bruxelles, 1050 Brussels, Belgium}

\author{\u{C}aslav Brukner}
\affiliation{Institute for Quantum Optics and Quantum Information (IQOQI-Vienna) of the Austrian Academy of Sciences, Boltzmanngasse 3, A-1090, Vienna, Austria}
\affiliation{Faculty of Physics, University of Vienna, Boltzmanngasse 5, A-1090 Vienna, Austria}

\maketitle

\begin{abstract}
  The quantum measurement problem can be regarded as the tension between the two alternative dynamics prescribed by quantum mechanics: the unitary evolution of the wave function and the state-update rule (or ``collapse'') at the instant a measurement takes place. The notorious Wigner's friend gedankenexperiment constitutes the paradoxical scenario in which different observers (one of whom is observed by the other) describe one and the same interaction differently, one --the Friend-- via state-update and the other --Wigner-- unitarily. This can lead to Wigner and his friend assigning different probabilities to the outcome of the same subsequent measurement. In this paper, we apply the Page-Wootters mechanism (PWM) as a timeless description of Wigner's friend-like scenarios. We show that the standard rules to assign two-time conditional probabilities within the PWM need to be modified to deal with the Wigner's friend gedankenexperiment. We identify three main definitions of such modified rules to assign two-time conditional probabilities, all of which reduce to standard quantum theory for non-Wigner's friend scenarios. However, when applied to the Wigner's friend setup each rule assigns different conditional probabilities, potentially resolving the probability-assignment paradox in a different manner. Moreover, one rule imposes strict limits on when a joint probability distribution for the measurement outcomes of Wigner and his Friend is well-defined, which single out those cases where Wigner's measurement does not disturb the Friend's memory and such a probability has an operational meaning in terms of collectible statistics. Interestingly, the same limits guarantee that said measurement outcomes fulfill the consistency condition of the consistent histories framework.
\end{abstract}

\section{Introduction}
\label{Introduction}

Standard quantum theory features two dynamical processes: the so-called ``collapse of the wave function''  that occurs during a measurement and the unitary evolution describing the propagation of the wave function in the absence of measurements. However, quantum theory itself does not specify when each process should apply, leading to the well-known quantum measurement problem\footnote{In the vast literature devoted to this topic, the ``quantum measurement problem'' does not have a unique, clear definition. Intuitively, it can be thought of as the problem of \emph{how}, \emph{when} and \emph{under what circumstances} definite values of physical variables are obtained~\cite{busch1996quantum}.
However, in this paper we focus on the ambiguity of the application between unitary and ``collapse'' dynamics.}  \cite{maudlin1995three,bub2010two}.

This situation is illustrated by the gedankenexperiment of Wigner's friend~\cite{wignerRemarksMindBodyQuestion1995}. An observer $F$ (the Friend) performs a measurement on a quantum system $S$ in a closed laboratory, the outcome of which is recorded by the click of a measuring apparatus and/or as a definite record in the Friend's memory inside the laboratory. After the  measurement, $F$ assigns a state to $S$ conditioned on her measurement outcome (i.e., state update). Simultaneously, another observer $W$ (Wigner) situated outside of the laboratory, would describe the entire measurement by $F$ as unitary, resulting in him assigning a specific  entangled state to $F$ and $S$. This leads to the paradoxical situation that the two observers $F$ and $W$ assign different states after $F$'s measurement and, hence, different probabilities to the outcomes of subsequent measurements.
As shown in Refs.~\cite{brukner2017quantum,frauchigerQuantumTheoryCannot2018,bruknerNoGoTheoremObserverIndependent2018},
the seemingly natural assumptions that quantum theory is universal, that different observers can each apply either of the two dynamical processes of the theory with respect to their description, and that the reasoning of different agents about one another can be freely combined to make statements about an objective reality lead to contradictions for Wigner's friend setups. These contradictions can be regarded as resulting from an observer dependent application of the state-update rule and involve inferences about the outcomes different observers obtain at different times.\\

We begin by formulating the Wigner's friend gedankenexperiment in terms of the a priori  timeless formulation of quantum theory put forward by Page and Wootters~\cite{pageEvolutionEvolutionDynamics1983}. In such an approach time evolution emerges from quantum correlations between a clock and a system whose dynamics the clock will track. {The advantage of the timeless approach in the context of the Wigner's friend gedankenexperiment is that it assigns a timeless state from which one can compute  probabilities of an event conditioned on another, regardless of their temporal order as indicated by the clock.}
This allows us to {ask} the following two questions: 
\begin{enumerate}
\item What is the probability that $W$ measures outcome $w$ at clock time $t_2$, given that $F$ has measured outcome $f$ at a previous clock time $t_1$?
\item What is the probability that $F$ measures outcome $f$ at clock time $t_1$, conditioned on the fact that $W$ will measure outcome $w$ at a later clock time $t_2$?
\end{enumerate}

We will show that there exist more than one way to consistently assign conditional probabilities using the Page-Wootters mechanism (PWM), all of which reduce to standard quantum conditional probabilities for non-Wigner’s friend scenarios. Depending on the choice of conditional-probability rule, however, the Wigner’s friend gedankenexperiment leads to different probabilities for the results of Wigner and his Friend, thus resolving the paradox of the ambiguous probability assignments in different manners. {Using the PWM all observers will agree on the overall state and assuming they all use the same probability rule they will agree in their probability assignments.}
The first rule gives probabilities that correspond to applying the state-update rule after the Friend's measurement and, hence,  suggests collapse dynamics for all observers. The other two rules are in accordance with unitary evolution
and give conditional probabilities different from the first rule. Moreover, they give well-defined probabilities only under certain conditions implying that the above two questions can only be meaningfully answered for specific settings. 
In case of the last rule, these conditions single out those settings for which Wigner's measurement does not disturb the Freind's memory and, hence, the respective probabilities correspond to collectible statistics for the Friend.

\section{The Page and Wootters Mechanism}
\label{The Page and Wootters Mechanism}

We now summarize the Page-Wootters mechanism~\cite{pageEvolutionEvolutionDynamics1983,woottersTimeReplacedQuantum1984}, which begins by considering a clock and system whose dynamics the clock tracks. The clock and system are described respectively by the Hilbert spaces $\mathcal{H}_C \simeq L^2(\mathbb{R})$ and $\mathcal{H}_S$. The joint state $\kket{\Psi}$ assigned to the clock and system is a solution to a Wheeler-DeWitt-like equation
\begin{align}
\hat{H }\ket{\Psi}\rangle = \left(\hat{H}_C +\hat{H}_S\right) \ket{\Psi}\rangle =  0, \label{WheelerDeWitt}
\end{align}
where $\hat{H}_S$ is the system Hamiltonian and $H_C =\hat{P}_C$ is the clock Hamiltonian, which is taken to be the momentum operator $\hat{P}_C$ on $\mathcal{H}_C$.
In general, a Wheeler-DeWitt-like equation can in some cases be interpreted as arising from the canonical quantization of a gauge theory with a Hamiltonian constraint.  General relativity is an example of such a gauge theory and it is for this reason that Page and Wootters first put forward their formulation of quantum theory to address the problem of time~\cite{kucharTimeInterpretationsQuantum2011,Isham1993}.
 Solutions to Eq.~\eqref{WheelerDeWitt} can be obtained by acting on states in the \emph{kinematical} Hilbert space $\mathcal{K} \simeq \mathcal{H}_C \otimes \mathcal{H}_S$ with the  operator 
\begin{align}
P^{\rm ph} \ce \int_{\mathbb{R}} ds \, e^{-is\hat{H}}, 
\label{Pphys}
\end{align}
that is, supposing $\ket{\psi} \in \mathcal{K}$, then  $\kket{\Psi} = P^{\rm ph} \ket{\psi}$ is a solution to Eq.~\eqref{WheelerDeWitt}. Physical states of the theory, $\kket{\Psi} \in \mathcal{H}$, are elements of the \emph{physical} space $\mathcal{H}$, which is not a subspace of the kinematical Hilbert space $\mathcal{K}$. This is because the spectrum of $\hat{H}$ is continuous around zero, from which it follows that physical states are not normalizable in the kinematical inner product~\cite{rovelliQuantumGravity2004,kieferQuantumGravity2012}. Thus a new inner product must be used to normalize the physical states, which in turn defines the physical Hilbert space $\mathcal{H}$; such an inner product is defined in Eq.~\eqref{PhysicalInnerProduct} and motivated in what follows. 

First, one defines the time read by the clock as the measurement outcome of a time observable $\hat{T}_C$ that is covariant~\cite{holevoProbabilisticStatisticalAspects1982,buschOperationalQuantumPhysics} with respect to the clock Hamiltonian~$\hat{H}_C$. By covariant it is meant that supposing the spectral decomposition of the time observable on $\mathcal{H}_C$ is
\begin{align}
\hat{T}_C = \int_{\mathbb{R}} dt\,  t \ket{t}\!\bra{t},
\end{align}
where $\ket{t}$ are eigenkets of $\hat{T}_C$ associated with the eigenvalue $t \in \mathbb{R}$, the eigenkets are connected to one another by the unitary generated by the clock Hamiltonian, $\ket{t'} = e^{-i \hat{H}_C (t'-t)} \ket{t}$. In this case, enforcing the covariance condition implies that $\hat{T}_C$ is canonically conjugate to the clock Hamiltonian, $[\hat{T}_C , \hat{H}_C] = i $, and thus is equivalent to the position operator on $\mathcal{H}_C$.

Now consider a system observable $\hat{M} = \sum_{m} m \Pi_m$, where $\{\Pi_m, \ \forall \, m \in \spec{\hat{M}}\}$ defines a projective valued measure on $\mathcal{H}_S$. The probability that $\hat{M}$ takes the value $m$ when the clock reads time $t$ is given by the Born rule
\begin{align}
&P\left(\hat{M} = m  \ \mbox{when}\  \hat{T}_C = t \right) \nn \\
&\qquad = \frac{\bbra{\Psi}  \big(\proj{t} \ox \Pi_m   \big) \kket{\Psi}}{\bbra{\Psi} \big(\proj{t} \ox \mathds{1}_S \big)  \kket{\Psi}}.
\label{AwhenT}
\end{align}
In what follows, we will use an abbreviated expression to refer to these probabilities in which we omit the operators, i.e., $P(\hat{M} = m  \ \mbox{when}\  \hat{T}_C = t ) \equiv P( m  \ \mbox{when}\  t )$.

The form of Eq.~\eqref{AwhenT} suggests defining the conditional state of the system given the clock reads the time $t$ as
\begin{align}
\ket{\psi_S(t)} \ce \big( \bra{t} \otimes \mathds{1}_S \big) \kket{\Psi}.
\label{ConditionalState}
\end{align}
Further, demanding that the conditional state remains normalized for all $t$ implies that the physical state $\kket{\Psi}$ must be normalized with respect to the physical inner product~\cite{smithQuantizingTimeInteracting2019}
\begin{align}
\braket{\braket{\Psi | \Psi}}_{\rm PW} \ce \braket{\braket{\Psi | \big(\proj{t} \ox \mathds{1}_S \big) |\Psi }} = 1,
\label{PhysicalInnerProduct}
\end{align}
for all $t \in \mathbb{R}$, from which it follows that $\braket{\psi_S(t) | \psi_S(t)} = 1$. Given Eqs.~ \eqref{ConditionalState} and \eqref{PhysicalInnerProduct}, the probability in Eq.~\eqref{AwhenT} may be expressed as
\begin{align}
P\left( m  \ \mbox{when}\  t \right) &=  \braket{\psi_S(t)| \Pi_m  |\psi_S(t)}.
\label{AwhenT2}
\end{align}
Since Eq.~\eqref{AwhenT2} is identical to the probability assigned to the measurement outcome $m$ given the state of the system is $\ket{\psi_S(t)}$ by the Born rule in the standard formulation of quantum theory, we are justified in identifying the conditional state in Eq.~\eqref{ConditionalState} as the standard time-dependent wave function. Further, the covariance condition satisfied by the eigenstates of $\hat{T}_C$ implies that the conditional state satisfies the Schr\"{o}dinger equation~\cite{woottersTimeReplacedQuantum1984}.

We note that because the eigenstates of $\hat{T}_C$ can be used to form a resolution of the identity on the clock Hilbert space, $\mathds{1}_C = \int dt \, \proj{t}$, physical states may be expressed as
\begin{align}
\kket{\Psi} = \int_{\mathbb{R}} dt \, \ket{t} \ket{\psi_S(t)}.
\end{align}
From the above equation one can immediately see that in general physical states describe an entangled state of the clock and the system. It is the correlations between the clock and system described by this entangled state that are responsible for the relative evolution of the system with respect to the clock.\\

\section{Timeless formulation of the Wigner's friend experiment}
\label{purifiedM}

\begin{figure}[t]
\centering
\includegraphics[width= 0.46\textwidth]{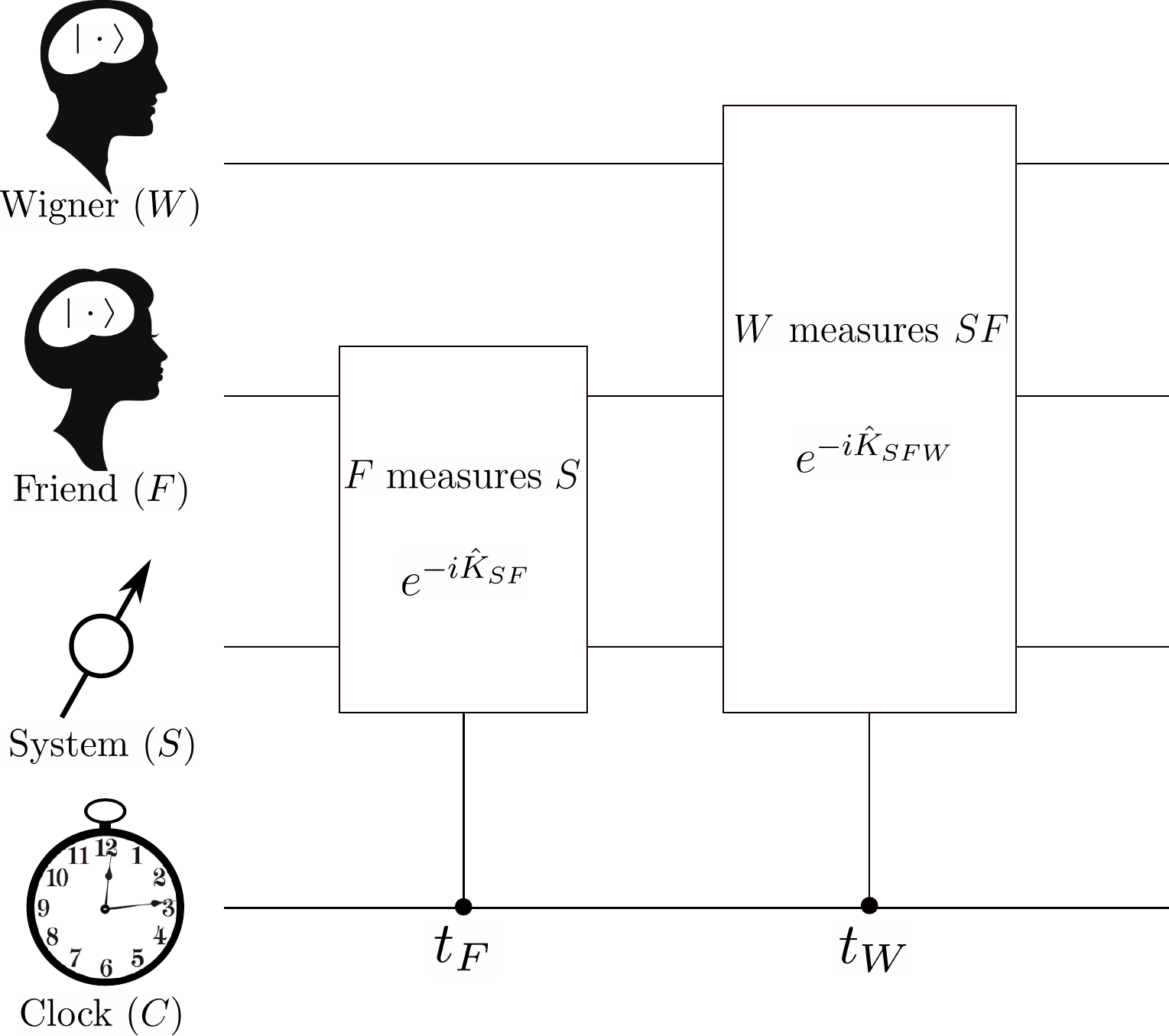}
\caption{The circuit representation of the Wigner’s friend experiment as encoded in the physical state $\kket{\Psi}$ given in Eq.~\eqref{historystate}. This can be thought of as the perspective of a hypothetical higher level observer whose experimental powers are even beyond that of Wigner. As described in the main text, at time $t_F$ according to clock $C$ the Friend performs a measurement and her memory gets entangled with the system $S$. At another time $t_W>t_F$ Wigner measures both $S$ and $F$ and his memory gets entangled with the joint system.}\label{CircuitDiagram}
\label{circuit}
\end{figure}

We now present a description of the Wigner's friend experiment depicted in Fig.~\ref{CircuitDiagram} in terms of the Page and Wootters mechanism. 

Consider a two-level system ($S$) associated with the Hilbert space $\mathcal{H}_S \simeq \mathbb{C}^2$ and a projective valued measurement of $S$ that yields two possible outcomes, $\{\uparrow, \downarrow\}$. Such a measurement is described by the set of projectors $\{\Pi_\uparrow, \Pi_\downarrow\} \subset \mathcal{E}(\mathcal{H}_S)$, where $\Pi_\uparrow+ \Pi_\downarrow = \mathds{1}_S$ and $\mathcal{E}(\mathcal{H})$ denotes the space of effect operators on the Hilbert space $\mathcal{H}$. The Friend ($F$) makes such a measurement of $S$ at the time $t_F$ and the outcome is recorded by $F$'s measuring device and eventually in her memory, which we associate with the Hilbert space $\mathcal{H}_F \simeq \mathbb{C}^3$. Situated outside of $F$'s laboratory is Wigner ($W$), who makes a joint measurement of $S$ and $F$ with two possible outcomes, $\{\mbox{yes}, \mbox{no} \}$, at a time $t_W > t_F$. This measurement corresponds to the set of projectors $\{\Pi_{\rm yes}, \Pi_{\rm no}\} \subset \mathcal{E}(\mathcal{H}_S \otimes \mathcal{H}_F)$ with $\Pi_{\rm yes}+ \Pi_{\rm no}=\mathds{1}$, and its result is recorded in $W$'s memory, the state of which is encoded in the Hilbert space $\mathcal{H}_W \simeq \mathbb{C}^3$. The times $t_F$ and $t_W$ are associated with a clock system $C$ described by the Hilbert space $H_C \simeq L^2(\mathbb{R})$ as discussed in Sec.~\ref{The Page and Wootters Mechanism}.

Following Hellmann \emph{et al.} \cite{hellmann2007multiple} and Giovannetti  \emph{et al.} \cite{giovannetti2015quantum}, we use a von Neumann  measurement model to describe $F$'s and $W$'s measurement within the PWM. The physical states describing the Wigner's friend experiment in Fig.~\ref{CircuitDiagram} are solutions to
\begin{align}
\left(\! \hat{H}_C \!+\! \delta(\hat T \!- \!t_F) \hat{K}_{SF} \!+\! \delta(\hat T \! -\! t_W) \hat{K}_{SFW} \! \right) \!\kket{\Psi} \!=\! 0,
\label{HWigner}
\end{align}
where $\hat{K}_{SF}$ and $\hat{K}_{SFW}$ are the interaction Hamiltonians coupling respectively $S$ and $F$ during $F$'s measurement of $S$ and $S$, $F$, and $W$ during $W$'s measurement of $SF$:
\begin{align*}
&e^{-i\hat{K}_{SF}} \ket{\psi_S}\ket{R_F} = \sum_{f \in \{\uparrow, \downarrow\}} \Pi_f\ket{\psi_S}\ket {f_F}, \\
& e^{-i\hat{K}_{SFW}} \ket{\phi_{SF}}\ket{R_W} =\sum_{w \in \{{\rm yes}, \, {\rm no}\}} \Pi_w \ket {\phi_{SF}}\ket{w_W},
\end{align*}
where $\ket{\psi_S}$ is the state of the system for clock readings $t<t_F$ and $\ket{R_F} \in \mathcal{H}_F$, $\ket{R_W} \in \mathcal{H}_W$ are the pre-measurement ``ready'' states of the memories of $F$ and $W$. The sets $\{\ket{R_F}, \ket{\uparrow_F}, \ket{\downarrow_F}\}$ and $\{\ket{R_W}, \ket{{\rm yes}_W}, \ket{{\rm no}_W}\}$ form  orthonormal bases for $\mathcal{H}_F$  and $\mathcal{H}_W$ respectively. Using this measurement scheme we will calculate the probabilities of result $m$ by projecting onto the memory entries of the respective observer. Hence, the formula  in Eq.~\eqref{AwhenT} for arbitrary time $t$ becomes
\begin{align}
P\left(m  \ \mbox{when}\   t \right) &= \frac{\bbra{\Psi}  \big(\proj{t} \ox \Pi^m   \big) \kket{\Psi}}{\bbra{\Psi} \big(\proj{t} \ox \mathds{1} \big)  \kket{\Psi}},
\label{AwhenTpure}
\end{align}
where $\Pi^m$ now acts on $\mathcal{H}_M$, the Hilbert space of the apparatus or memory storing the result, and $\kket{\Psi}$ is a solution to Eq.~\eqref{HWigner}. 
For simplicity we have assumed no free dynamics of $S$, $F$, and $W$ in Eq.~\eqref{HWigner} in addition to their interaction with one another during the measurements. The solutions take the form
\begin{align}
\ket\Psi\rangle &= \int_{-\infty}^{t_F}dt \,\ket t\ket{\psi_S}\ket {R_F}\ket{R_W} \nn \\
&\quad +  \int_{t_F}^{t_W}dt \, \ket t \sum_{f \in \{\uparrow, \downarrow\}} \Pi_f\ket{\psi_S}\ket {f_F}\ket{R_W}  \nn\\
&\quad +  \int_{t_W}^{\infty}dt\, \ket t  \! \! \! \sum_{\substack{f \in \{\uparrow, \, \downarrow\} \\ w \in \{{\rm yes}, \, {\rm no}\}}} \! \! \! \Pi_w \Pi_f \ket{\psi_S}\ket {f_F}\ket{w_W} \!,
\label{historystate}
\end{align}
for arbitrary states $\ket{\psi_S}$ of the system. When expressed in the clock basis like in Eq.~\eqref{historystate} the physical state encodes the complete history of system $S$ and the measurements performed on it. In what follows we will explicitly consider states
\begin{align}
\ket{\psi_S} = a \ket{\uparrow_S} + b e^{i\phi_S}\ket{\downarrow_S},
\end{align}
where $a,b, \phi_{S} \in \mathbb{R}$ and
\begin{align}
\ket{ {\rm yes}_{SF}} = \alpha \ket{\uparrow_S} \ket{\uparrow_F} + \beta e^{i \phi_{SF}}\ket{\downarrow_S}\ket{\downarrow_F},
\label{MWigner}
\end{align}
where $\alpha,\beta, \phi_{SF} \in \mathbb{R}$, with $\Pi_{\rm yes} = \proj{\rm yes}$ and $\Pi_{\rm no} = \mathds{1}- \proj{\rm yes}$, moreover, we will use $\phi\ce \phi_S - \phi_{SF}$. Note that a measurement by Wigner as implied by Eq.~\eqref{MWigner} will in general affect the Friend's memory state and potentially alter her perceived result.

\section{Different definitions of conditional probabilities, different solutions of the Wigner's friend paradox}
\label{Sect:definitions}
We want to express the dynamical content of a theory in a manifestly operational way, that is, in terms of probabilities of measurement outcomes. As emphasized by Kucha\v{r}~\cite{kucharTimeInterpretationsQuantum2011}, \emph{the fundamental question in any dynamical theory} is: if one measures an observable at the time $t_1$ and obtains outcome $a$, what is the probability of a different measurement yielding the outcome $b$ at the time $t_2>t_1$?
For non-Wigner's friend scenarios there have been two proposals for defining probability rules within the Page-Wootters mechanism~\cite{dolby2004conditional,giovannetti2015quantum} in order to answer this fundamental question. Both approaches recover the standard quantum formalism for the situation of two observables measured at different times inquired above.

In order to deal with Wigner's friend scenarios (in particular with the setup depicted in Fig.~\ref{circuit}), however, one needs to generalize the standard definition of the conditional-probability rule. The rationale behind this generalization is that the state of the Friend's memory is --except for particular cases-- modified by Wigner's measurement. Hence, it is not any longer possible to simply read out the information in $F$'s memory at the end of the protocol in order to learn what outcome she obtained. Thus, we will be interested in calculating two-time  conditional probabilities at times $t_1$ (in between $F$'s and $W$'s measurement) and $t_2$ (after $W$'s measurement), i.e., $t_F<t_1<t_W<t_2$. We propose three main alternative conditional probability definitions, two of which are inspired by Refs.~\cite{dolby2004conditional} and~\cite{giovannetti2015quantum} respectively, as well as a third novel one. While all of these definitions give the standard quantum probabilities for non-Wigner's friend cases (as shown explicitly in Appendix~\ref{A:unitarity}), they give different conditional probabilities for the measurement results of Wigner and his Friend. 

First we propose a conditional-probability rule similar to that put forward by Dolby~\cite{dolby2004conditional}, where measurement operators $\proj{t_1}\ox \Pi_m $ together with the operator $P^{\rm ph}$ (defined in Eq. \eqref{Pphys}) are used to calculate two-time conditional probabilities. Specifically, after the application of one operator $\proj{t_1} \ox \Pi_m$ to the physical state $\kket{\Psi}$ the constraint in Eq.~\eqref{WheelerDeWitt} is in general no longer fulfilled, i.e., $\hat{H}( \proj{t_1}\ox\Pi_m \kket{\Psi})\neq0$. As proposed in~\cite{dolby2004conditional} one uses the operator $P^{\rm ph}$ to obtain again a physical state, $\hat{H}\big(P^{\rm ph} \proj{t_1}\ox\Pi_m \kket{\Psi}\big)=0$, before applying a second measurement operator $ \proj{t_2}\ox \Pi_n$.
 In order to describe Wigner's friend scenarios we account for the measurements as in Eqs.~\eqref{HWigner}-\eqref{historystate}.
\begin{widetext}
\onecolumn
\begin{description}
\item[Definition 1]
\textbf{(two-time ``collapse'').}
The conditional probability of result $n$ at time $t_2$ given result $m$ at time $t_1$ is
\begin{align} 
&P_1\left(  n \textrm{ when } t_2 \, |\, m \textrm{ when } t_1 \right)= \frac{\bbra{\Psi}\proj{t_1} \! \otimes \! \Pi^m P^{\rm ph } \proj{t_2}\otimes \Pi^n P\ph \ket{t_1}\!\bra{t_1} \!\otimes \!\Pi^m|\Psi\rangle\rangle}{\langle\langle \Psi| \ket{t_1}\!\bra{t_1} \!\otimes\! \Pi^m |\Psi\rangle\rangle},
\label{DefDolby}
\end{align}
where $ \Pi^m$ and $ \Pi^n$ are now the projectors on the respective states of the apparatus or memories in $\mathcal{H}_M$ and $\mathcal{H}_N$, and $P\ph$ was defined in Eq.~\eqref{Pphys}. 
\end{description}
\end{widetext}

Intuitively, the numerator in Eq.~\eqref{DefDolby}, in the fashion of Ref.~\cite{dolby2004conditional},  is the modulus square of the physical state $\kket{\Psi}$ which is first projected on $\Pi^m$ at time $t_1$, then brought back into the physical space by applying $P^{\rm ph }$ and finally projected on $\Pi^n$ at time $t_2$.
Definition~1 always gives well-defined probabilities, and for the results of Wigner and his Friend this leads to the conditional probabilities displayed in Table~\ref{ProbDef1}. We label this definition ``two-time collapse'' because any observer will agree on these probabilities assignments which correspond to applying the state-update rule (``collapse'') after every measurement. Moreover, it is genuinely two-time, since the expression explicitly depends on both the times $t_1$ and $t_2$. Hence, this resolves the probabilistic paradox for Wigner's friend scenarios insofar as it suggests that both Wigner and the Friend should describe $F$'s measurement with  ``collapse'' dynamics.
Note, however, that this rule does not give rise to a well-defined joint probability since $P_1\left(  n \textrm{ when } t_2 \, |\, m \textrm{ when } t_1 \right)\cdot P(m\textrm{ when } t_1) \neq P_1\left(  m \textrm{ when } t_1 \, |\, n \textrm{ when } t_2 \right)\cdot P(n\textrm{ when } t_2)$. This reflects the fact that the numerator in Eq. \eqref{DefDolby} is not invariant under the exchange of $ \proj{t_1}\otimes  \Pi^m$ and $\proj{t_2} \otimes \Pi^n$.\\

\begin{table}
\centering
\setlength{\tabcolsep}{30pt}
\begin{tabular}{c|c|c}
\multicolumn{3}{c}{
$\bm{P_1\left(  w \ {\rm when}\ t_2 \, |\, f \ {\rm when} \ t_1 \right) } $
}  \\ \hline \hline
\diagbox[innerwidth=20pt]{$f$}{$w$} & $\rm{yes}$ & $\rm{no}$ \\ \hline 
$\ua$ & $\alpha^2$ &$\beta^2$  \\
$\da$ & $\beta^2$ & $\alpha^2$  \\
\multicolumn{3}{c}{}\\
\multicolumn{3}{c}{
$\bm{P_1\left( f \ {\rm when} \ t_1  \, |\, w \ {\rm when}\ t_2\right) } $
}  \\ \hline \hline
\diagbox[innerwidth=20pt]{$f$}{$w$} & $\rm{yes}$ & $\rm{no}$ \\ \hline 
$\ua$ & $\alpha^2$ &$\beta^2$  \\
$\da$ & $\beta^2$ & $\alpha^2$  \\
\end{tabular}
\caption{The conditional probabilities of Wigner seeing result $w$ at time $t_2$ given that the Friend saw result $f$ at time $t_1$ and of the Friend seeing $f$ at $t_1$ given that Wigner will see $w$ at $t_2$ according to definition 1. Note that the two conditional probabilities are equal and hence, applying Bayes' rule will in general \emph{not} give rise to \emph{one} joint probability expression.}
\label{ProbDef1}
\end{table}

Alternatively, we consider the conditional probability rule proposed by Giovanetti \emph{et al.}~\cite{giovannetti2015quantum} in the context of Wigner's friend experiments. The authors of~\cite{giovannetti2015quantum} include generalized measurements as described in Eq. \eqref{HWigner}-\eqref{historystate} in the constraint Hamiltonian and apply projection operators on the measurement apparatus or memories at some time $t$ after both measurements in question have been performed, i.e., $\proj{t}\ox\Pi^m \ox \Pi^n$. This equates measurements at different times with one joint measurement of the records of the obtained results at a final single time, which is no longer a trivial thing to do when considering Wigner's friend scenarios.
The original formulation in~\cite{giovannetti2015quantum} states the following rule for calculating conditional probabilities.
\begin{description}
\item[Definition 2a]
\textbf{(two-time unitary, uninterpreted conditions).} The conditional probability (in the cases where it is well-defined) of result $n$ at time $t_2$ given result $m$ at time $t_1$ is
\begin{align}
&P_{2a}\left(  n \textrm{ when } t_2 \, |\, m \textrm{ when } t_1\right) \nn \\
&\qquad =\frac{\langle\langle \Psi \ket{t_2}\bra{t_2} \otimes \Pi^{n} \otimes \Pi^{m} \kket{\Psi}}{\langle\langle\Psi \ket{t_1}\bra{t_1} \ox \Pi^{m}\kket{\Psi}}.
\label{DefGio1}
\end{align}
\end{description}
The intuition behind this definition is to read out both $F$'s and $W$'s memories at the final time $t_2$ and take the modulus square to define a joint probability, and use that to construct a conditional probability. Namely, to take the expectation value of the operators $\Pi^m$ and $\Pi^n$ at time $t_2$ with the physical state $\kket{\Psi}$ (numerator). This is then divided by the one-time probability of $F$ finding outcome $m$ at time $t_1$ (denominator), i.e., before $W$'s measurement takes place. Equation~\eqref{DefGio1} depends on both measurement outcomes and on both times $t_1$ and $t_2$, thus amounting to a genuine two-time probability. However, as shown in Appendix~\ref{A:condiDef2}, in the Wigner's friend scenario of Fig.~\ref{circuit}, these expressions are not always proper probabilities, for they are normalized only if 
\begin{align}
&2\alpha^2\beta^2\left(\frac{b}{a}\right)^2+2\cos(\phi)(\alpha^3\beta-\alpha\beta^3)\frac{b}{a} \nn \\
&=2\alpha^2\beta^2\left(\frac{a}{b}\right)^2 -2\cos(\phi)(\alpha^3\beta-\alpha\beta^3)\frac{a}{b} \nn \\
&=1-\alpha^4-\beta^4.
\label{condiDef.2.a}
\end{align}
Under these conditions, definition 2a gives the probabilities stated in Table~\ref{ProbDef2.a}. While these expressions are formally probabilities when Eqs.~\eqref{condiDef.2.a} are satisfied, their operational meaning is admittedly not clear to us, except for special cases (see Sec.~\ref{outlook}).

\begin{table}
\setlength{\tabcolsep}{2pt}
\renewcommand{\arraystretch}{1.25}
\resizebox{.92\textwidth}{!}{\begin{minipage}{\textwidth}
\begin{tabular}{c|c|c}
\multicolumn{3}{c}{
$\bm{P_{2a}\left(  w \ {\rm when}\ t_2 \, |\, f \ {\rm when} \ t_1 \right) }$
}  \\ \hline \hline
\diagbox[innerwidth=20pt]{$f$}{$w$} & $\rm{yes}$ & $\rm{no}$ \\ \hline 
$\ua $ & $\frac{1 + 2 \alpha^2 + (\beta^4 -\alpha^4) \cos(2 \phi) \pm \chi }{4}$ & $\frac{1 + 2 \beta^2 + (\alpha^4-\beta^4) \cos(2 \phi) \mp \chi}{4}$ \\
$\da$ & $\frac{1 + 2 \beta^2 + (\alpha^4-\beta^4 ) \cos(2 \phi) \pm\chi }{4}$ & $\frac{1 + 2 \alpha^2 + (\beta^4 -\alpha^4) \cos(2 \phi) \mp \chi}{4}$  \\[5pt]
\multicolumn{3}{c}{with $\chi \ce 2 \cos(\phi) \sqrt{1 - (\alpha^2 - \beta^2)^2 \sin^2(\phi)}$}\\
\end{tabular}
\end{minipage}}
\caption{The conditional probabilities of Wigner seeing result $w$ at time $t_2$ given that the Friend saw result $f$ at time $t_1$ according to definition 2a. The different signs correspond to the different solutions of the quadratic equations in Eqs.\eqref{condiDef.2.a}. Note that there is no sensible definition of $P_{2a}\left( f \ {\rm when }\ t_1 \, |\, w \ {\rm when }\ t_2\right)$, since the numerator in Eq.\eqref{DefGio1} does not depend on $t_1$ and $\langle\langle{\Psi}  \ket{t_1}\bra{t_1}\Pi^{w}\ox \kket{\Psi}=0$.}
\label{ProbDef2.a}
\end{table}

Considering the general arguments in~\cite{hellmann2007multiple} in favor of equating measurements at different times with one collective measurement at a final time, we propose a straightforward modification of definition 2a for conditional  probabilities.
\begin{description}
\item[Definition 2b]
\textbf{(one-time unitary).}
The conditional probability of result $n$ given $m$ is
\begin{align}
&P_{2b}\left(  n \textrm{ when } t_2 \, |\, m \textrm{ when } t_2\right)  \nn \\
&\quad =  \frac{\langle\langle\Psi \ket{t_2}\bra{t_2} \ox \Pi^{n} \otimes \Pi^{m} \kket{\Psi}}{\langle\langle\Psi\ket{t_2}\bra{t_2}  \ox \Pi^{m}\kket{\Psi}},
\label{DefGio2}
\end{align}
where $t_2$ is some time after the second measurement has been performed.
\end{description}
These one-time probabilities are always well-defined and shown in Table~\ref{ProbDef2.b}. They correspond to jointly observable statistics after full unitary evolution (i.e., without state-update) and are equivalent to the proposal in \cite{baumann2018formalisms}. This resolves the paradox of the ambiguous probability assignment in Wigner's friend scenarios insofar as it corresponds to all observers describing all measurements unitarily up to that point where they can all compare their records, that is, the entries in $F$'s and $W$'s memories after both measurements took place. {It is important to note that $F$'s memory will in general be altered by $W$'s measurement and reading it out at the end will no longer correspond to what was encoded after $F$'s measurement.}
Note that for non-Wigner's friend scenarios definitions 2a and 2b are equivalent, since in this case $\langle\langle{\Psi} \ket{t_2}\bra{t_2} \ox\Pi^{m}\kket{\Psi}=\langle\langle{\Psi} \ket{t_1}\bra{t_1} \ox\Pi^{m} \kket{\Psi}$ (see Appendix~\ref{A:unitarity}). This is due to the fact that the memory, in which the result $m$ is stored at some time $t_1$, undergoes no further evolution until time $t_2$ and after. \\

\begin{table}
\setlength{\tabcolsep}{12pt}
\renewcommand{\arraystretch}{1.25}
\resizebox{.92\textwidth}{!}{\begin{minipage}{\textwidth}
\begin{tabular}{c|c|c}
\multicolumn{3}{c}{
$\bm{P_{2b}\left(  w \ {\rm when}\ t_2 \, |\, f \ {\rm when} \ t_2 \right) }$
}  \\ \hline \hline
\diagbox[innerwidth=20pt]{$f$}{$w$} & $\rm{yes}$ & $\rm{no}$ \\ \hline 
$\ua$ & $\frac{\frac{\alpha^2}{\beta^2}+2\frac{b\alpha}{a\beta}\cos(\phi)+\frac{b^2}{a^2}}{N_{\ua}}$ &$\frac{\frac{\beta^2}{\alpha^2}-2\frac{b\beta}{a\alpha}\cos(\phi)+\frac{b^2}{a^2}}{N_{\ua}}$ \\
$\da$ & $\frac{\frac{\beta^2}{\alpha^2}+2\frac{a\beta}{b\alpha}\cos(\phi)+\frac{a^2}{b^2}}{N_{\da}}$ & $\frac{\frac{\alpha^2}{\beta^2}+2\frac{a\alpha}{b\beta}\cos(\phi)+\frac{a^2}{b^2}}{N_{\da}}$  \\[5pt]
\multicolumn{3}{c}{with $N_{\ua}\ce \frac{\alpha^2}{\beta^2}+\frac{\beta^2}{\alpha^2}+2\frac{b}{a}\cos(\phi)\left( \frac{\alpha}{\beta}-\frac{\beta}{\alpha}\right)+2\frac{b^2}{a^2}$}\\
\multicolumn{3}{c}{and $N_{\da}\ce \frac{\alpha^2}{\beta^2}+\frac{\beta^2}{\alpha^2}+2\frac{a}{b}\cos(\phi)\left(\frac{\beta}{\alpha}- \frac{\alpha}{\beta}\right)+2\frac{a^2}{b^2}$}\\
\multicolumn{3}{c}{}\\
\multicolumn{3}{c}{
$\bm{P_{2b}\left( f \ {\rm when} \ t_2  \, |\, w \ {\rm when}\ t_2\right) } $
}  \\ \hline \hline
\diagbox[innerwidth=20pt]{$f$}{$w$} & $\rm{yes}$ & $\rm{no}$ \\ \hline 
$\ua$ & $\alpha^2$ &$\beta^2$  \\
$\da$ & $\beta^2$ & $\alpha^2$  \\
\end{tabular}
\end{minipage}}
\caption{The conditional probabilities for results $w$ of Wigner and $f$ of the Friend according to definition~2b. The joint probability is well-defined and corresponds to the numerator in Eq.~\eqref{DefGio2}. }
\label{ProbDef2.b}
\end{table}

The fact that the Friend's memory is the \emph{only} record of her observed result and is in general affected by Wigner's measurement raises the question whether it is operationally meaningful to assign a joint probability to the results $f$ and $w$. By construction of the setup these two results are in general not jointly observable by any single observer. However, as discussed in Ref.~\cite{baumannWignerFriendRational2019a}, in the special case of the joint state of the quantum system and the Friend being in an eigenstate of Wigner's measurement, $F$'s memory is not affected (a \emph{non-disturbance measurement}) and she can, in principle, learn both results. In fact, in this case Wigner can send his observed outcomes to the Friend who can combine them with her own (left undisturbed by Wigner's measurement) and construct the joint probability distribution.
We, thus, propose a third rule for calculating conditional two-time probabilities, which agrees with standard quantum theory for non-Wigner's friend scenarios and gives well-defined probabilities for Wigner's friend setups, in the cases where these probabilities are operationally meaningful.
\begin{description}
\item[Definition 3] 
\textbf{(two-time unitary, consistency conditions).}
The conditional probability (in the cases where it is well-defined) of result $n$ at time $t_2$ given result $m$ at time $t_1$ is 
\begin{align} 
&P_3\left(  n \textrm{ when } t_2 \, |\, m \textrm{ when } t_1 \right) = \nn \\
&\frac{\langle\langle\Psi  |(\ket{t_2} \bra{t_2}\otimes \Pi^{n}) P^{\rm ph}  (\ket{t_1}\bra{t_1}\otimes\Pi^m) \kket{\Psi}}{ \langle\langle\Psi  \ket{t_1}\bra{t_1}\otimes\Pi^m  \kket{\Psi}}.
\label{DefUs}
\end{align}
\end{description}
\begin{table}[t]
\setlength{\tabcolsep}{24pt}
\renewcommand{\arraystretch}{1.25}
\resizebox{.92\textwidth}{!}{\begin{minipage}{\textwidth}
\begin{tabular}{c|c|c}
\multicolumn{3}{c}{
$\bm{P_3\left(  w \ {\rm when}\ t_2 \, |\, f \ {\rm when} \ t_1 \right) } $
}  \\ \hline \hline
\diagbox[innerwidth=20pt]{$f$}{$w$} & $\rm{yes}$ & $\rm{no}$ \\ \hline 
$\ua$ & $1$ (or $0$) &$0$ (or $1$)  \\
$\da$ & $1$ (or $0$) & $0$ (or $1$)  \\
\multicolumn{3}{c}{}\\
\multicolumn{3}{c}{
$\bm{P_3\left( f \ {\rm when} \ t_1  \, |\, w \ {\rm when}\ t_2\right) } $
}  \\ \hline \hline
\diagbox[innerwidth=20pt]{$f$}{$w$} & $\rm{yes}$ & $\rm{no}$ \\ \hline 
$\ua$ & $\alpha^2$ (or $0$) & 0 (or $\beta^2$) \\
$\da$ & $\beta^2$  (or $0$) & 0 (or $\alpha^2$)  \\
\end{tabular}
\end{minipage}}
\caption{The conditional probabilities of Wigner seeing result $w$ at $t_2$ and the Friend seeing result $f$ at time $t_1$ (above) and of the Friend seeing $f$ at $t_1$ given that Wigner will see $w$ at $t_2$ (below), according to definition 3. the probabilities  are well-defined only in the case where Wigner's measurement is non-disturbing, i.e. $a=\alpha$ and $b=\beta$ (or $a=\beta$ and $b=-\alpha$).}
\label{ProbDef3}
\end{table}
This rule gives well-defined probabilities under the condition that the measurement operators commute on the physical state when compared at the same instant of time, i.e.
\begin{align} 
\left[\mathcal{U}(t_1,t_2)\Pi^m \mathcal{U}^\dagger(t_1,t_2), \Pi^n \right]	\kket{\Psi} = 0, 
\label{condition}
\end{align}
where $\mathcal{U}(t_2,t_1)=\bra{t_2}P^{\rm ph}\ket{t_1}$.\\ 
{We note, that condition \eqref{condition} ensures that $m$ and $n$ are part of a family of consistent quantum histories~\cite{griffiths2003consistent} as shown in Appendix~\ref{A:CH}. According to the consistent histories framework only within such a family can one meaningfully assign a set of probabilities to the properties encoded in the respective histories; in the case considered here, the results of Wigner and his Friend.
This agrees with the fact that definition 3 singles out the non-disturbing measurements by Wigner in the setup depicted in Fig.~\ref{circuit} (see Appendix~\ref{A:condiDef3} for explicit calculations), {namely
\begin{align} 
\phi= n\pi \qquad \text{ and } \qquad \frac{a}{\alpha}=\frac{b}{\beta}.
\label{condiDef3n.d.}
\end{align}}
The probabilities given by definition 3 are listed in Table~\ref{ProbDef3} and coincide with those given by definitions 2a and 2b (i.e., without the application of the state-update). In contrast to definition 2b, however, definition 3 describes genuine two-time probabilities, which, contrary to those given by definition 2a, have a clear interpretation and operational meaning.}  However, note that collecting these statistics is not straight-forward, for also in the non-disturbance case the Friend cannot simply send out her observed results to Wigner without changing the probabilities for Wigner's result (see Ref.~\cite{baumannWignerFriendRational2019a}). Yet, Wigner can in principle send inside the laboratory his measured data, from which the Friend can construct a joint (and conditional) probability distribution.

\section{Discussion and outlook}
\label{outlook}

We have presented three generalizations of the standard rule to assign probabilities to consecutive quantum measurements in a Wigner's friend scenario using the Page-Wootters mechanism. We showed how these probability rules potentially remove, in different manners, the ambiguity between the application of unitary dynamics and the state-update rule (``collapse''). While these represent a formal resolution (in fact a few distinct ones) of the probability-assignment paradox in Wigner's friend-like experiments, a number of key issues remain open: There are potentially more than just the three probability rules we presented, possibly leading to other resolutions of the paradox. Is there a way of classifying all of them? Furthermore, what is the operational meaning of each of the expressions that formally give probabilities?

Concerning the first question, we ought to stress that the only necessary condition that we have required from our definitions of two-time probabilities was, in fact, that they have to reduce to the standard Born rule in the non-Wigner's friend case, and that they give well-defined probabilities for at least certain cases of a Wigner's friend scenario. However, one can imagine further possible definitions that comply with these requirements, hence, a more systematic study would be desirable. {In particular, one extension of the original proposal of Page and Wootters put forward by Gambine \emph{et al}. is to compute conditional probabilities between averages over `evolving constants of motion',\footnote{More precisely, one would construct a family of Dirac observables, which commute with the constraint in Eq.~\eqref{WheelerDeWitt}, comprised of a partial observables associated with a clock and observables on Wigner's and his friend's memories. Such a family of Dirac observables is parametrized by the times read by the clock. Conditional probabilities would be computed as in Eq.~\eqref{AwhenT}, with the important difference that the numerator and denominator are averaged over all readings of the clock. See Ref.~\cite{gambiniConditionalProbabilitiesDirac2009} for more detail.} which would be interesting to examine in Wigner's friend setups.}

\begin{figure}[h!]
\centering
\includegraphics[width= 0.47\textwidth]{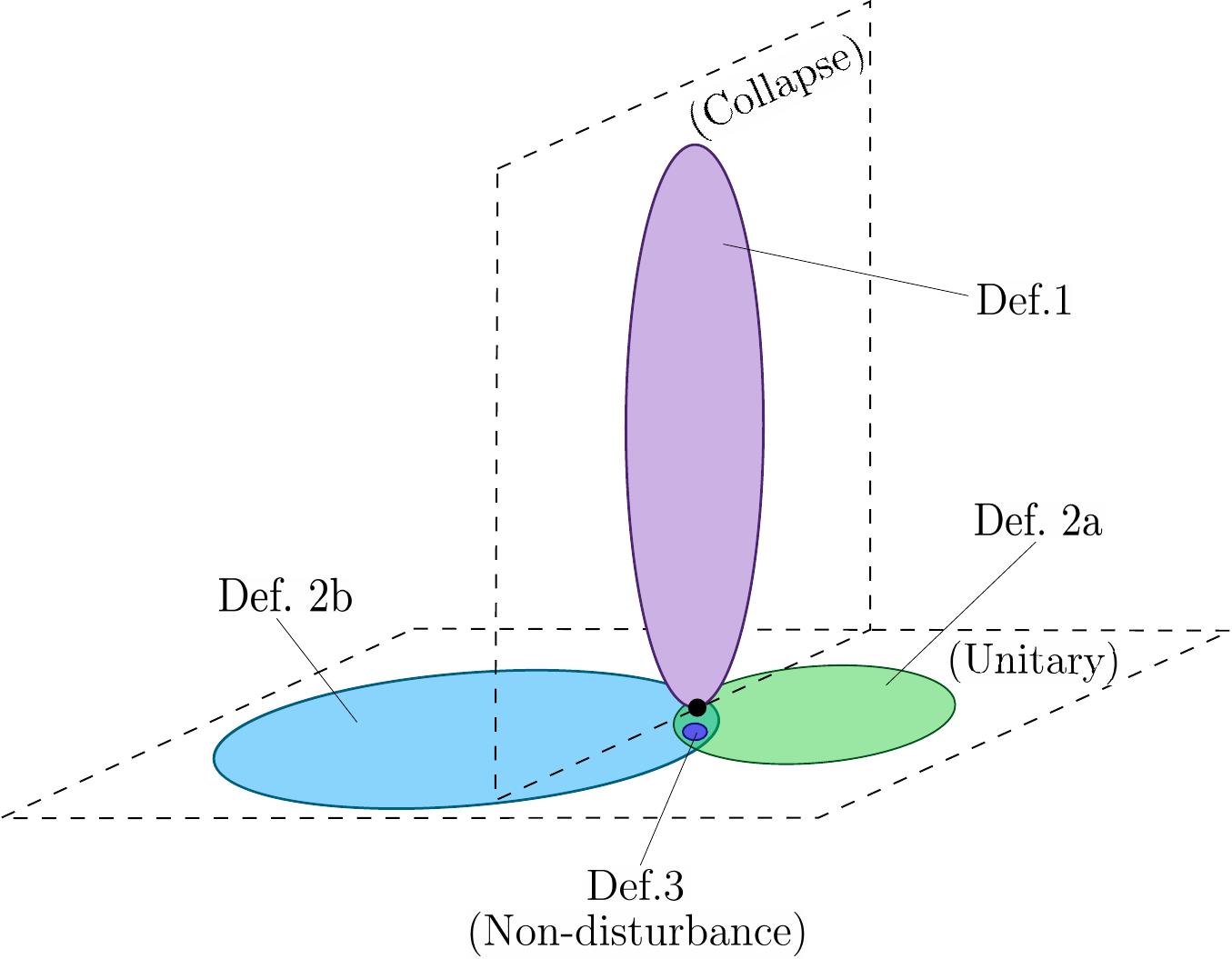}
\caption{Suggestive graphical representation of the probabilities according to the different definitions {in a Wigner's friend scenario (when they are well-defined). For non Wigner's friend scenarios all the definitions recover standard quantum probabilities, thus they fully overlap.}
The two planes distinguish between joint probabilities that are in accordance with applying (collapse) or not applying (unitary) the state-update rule after the Friend's measurement.
In the non-disturbance case (i.e., $\phi=n\pi$ and $a/\alpha=\pm b/\beta$) definition 2a, definition 2b and definition 3 give the same unitary, conditional probabilities, that are different from the collapse ones given by definition 1. There is, however, at least one instance where definitions 1, 2a, and 2b coincide (i.e., $\phi=(n+1/2)\pi$, $a=b=1/\sqrt2$ and arbitrary $\alpha$. 
\label{fig_conditions}}
\end{figure}
\begin{table*}[t]
\setlength{\tabcolsep}{5pt}
\renewcommand{\arraystretch}{2}
\resizebox{0.95\textwidth}{!}{\begin{minipage}{\textwidth}
\begin{tabular}{|m{1.25cm} |>{\centering} m{2cm}  | >{\centering}m{2cm} |>{\centering} m{2cm} | >{\centering}m{2cm} | >{\centering}m{2cm} | p{4cm}|}
\cline{1-7}
& standard QT & positive & normalized & two-time & sym. joint & Conditions\\
 \hline \hline
Def.~1 &\Large\ding{51} &\Large\ding{51} &\Large\ding{51} &\Large\ding{51} &\Large\ding{55}&\\
\hline
 & &  &  & &  & \\ [.5em]
Def.~2a & \Large\ding{51} &\Large\ding{51}&{\color{Bblue}\Large\ding{55}} &\Large\ding{51} &\Large\ding{51}&
\vspace{-4.75em}{\color{Bblue} normalized, if \newline
$\alpha^4+\beta^4+2\cos(\phi)(\alpha^3\beta-\alpha\beta^3)\frac{b}{a}+2\alpha^2\beta^2\left(\frac{b}{a}\right)^2=1$\newline
and \newline
$\alpha^4+\beta^4-2\cos(\phi)(\alpha^3\beta-\alpha\beta^3)\frac{a}{b}+2\alpha^2\beta^2\left(\frac{a}{b}\right)^2=1$.}
\\
\hline
Def.~2b & \Large\ding{51} &\Large\ding{51}  & \Large\ding{51} &\Large\ding{55} &\Large\ding{51} &\\
\hline
 & &  &  & &  & \\[-1.25em] 
Def.~3 & \Large\ding{51} & {\color{Bblue} \Large\ding{55}}&\Large\ding{51} & \Large\ding{51}&{\color{Bblue} \Large\ding{55}}& \vspace{-3em}{\color{Bblue} positive for non-disturbance measurement: $a=\alpha, b=\beta$ or $a=\beta, b=-\alpha$. And $\phi=n\pi$.}\\ 
\hline
\end{tabular}
\end{minipage}}
\begin{minipage}{\textwidth}
\caption{Comparison of the different proposed conditional probability rules for describing Wigner's friend experiments. From the left to the right, the columns indicate whether each definition: (i) reduces to standard quantum theory for non-Wigner's friend scenarios; (ii) is positive; (iii) is normalized; (iv) is a genuine two-time expression; (v) gives rise to a well-defined  joint probability distribution (i.e. $P(A)P(B|A)=P(B)P(A|B)$). Note, in fact, that definitions 2a and 3 become probabilities only under certain conditions, the blue \ding{55} indicates that the respective property is satisfied under certain conditions but not in general.
The last column on the right shows these conditions (if any) under which the respective properties are satisfied.}
\label{Tab:Comparison}
\end{minipage} 
\end{table*}
To address the second question, we now discuss the interpretation and the limits of each definition of conditional probabilities presented in the last section (see Table~\ref{Tab:Comparison} for a summary of the definitions and their properties, and Fig.~\ref{fig_conditions} for a sketched graphical representation). As already noticed, definition 1 has a clear interpretation in terms of ``objective collapse'' probabilities, but does not give rise to a well-defined joint probability (see Sec.~\ref{Sect:definitions}). However, one could maintain that at the operational level conditional probabilities come prior to joint probabilities, which are \textit{a posteriori} constructions. 

On the other hand, definition 2b gives measurement  statistics consistent with standard quantum theory assuming $S$, $F$ and $W$ evolve unitarily and the Born rule is applied after $W$'s measurement interaction has occurred. While this definition also has a clear interpretation, it fails in representing genuine two-time probabilities (only the final time $t_2$ appears in the rule to assign conditional probabilities for the measurements of $F$ and $W$).

Definition 3 has the merit of being interpretable as a genuine two-time probability, however it is a well-defined probability (i.e. real and non-negative) only under the rather demanding  ``non-disturbance measurement'' conditions, that is, when Wigner measures in a basis that contains the joint state of the system and the Friend (see Table~\ref{Tab:Comparison}). In this case, the probabilities are the same as definition 2b. The condition required  for definition 3 to result in genuine probabilities can be interpreted as  evidence that only in the non-disturbance case there exists a record of the past outcome $f$ of $F$'s measurement and the current outcome $w$ of $W$'s measurement. {Thus, only in this case is a joint probability distribution associated with the measurement outcomes of $W$ and $F$ operationally meaningful, since it corresponds to collectible statistics for the friend.} Remarkably, this limitation to the meaningful attribution of a joint probability of the outcomes of $F$ and $W$ emerges naturally from our formalism. This suggests a connection with a recent no-go theorem \cite{brukner2017quantum,bruknerNoGoTheoremObserverIndependent2018} stating that it is in general not possible to construct a joint probability distribution associated with measurement outcomes of different observers in a (more complicated) Wigner's friend scenario. Moreover, we have shown that the conditions for definition 3 to give well-defined probabilities ensure that the measurement outcomes of Wigner and his Friend are elements of a family of consistent quantum histories~\cite{griffiths2003consistent,losada2019frauchiger} (see Appendix~\ref{A:CH}).

Finally, definition 2a gives well-defined (i.e., normalized) probabilities only under the conditions in Eq.~\eqref{condiDef.2.a}. {Note that the non-disturbance measurement satisfies these conditions (see Fig.~\ref{fig_conditions}), meaning that if Eq.~\eqref{condiDef3n.d.} is satisfied, so is Eq.~\eqref{condiDef.2.a}.} In general, however, it is not clear to us how to interpret the probabilities given by definition 2a operationally. It seems that formally for any given initial state of the quantum system and fixed measurement of the Friend, there always exists a measurement choice of Wigner (besides the non-disturbance one) which ensures that definition 2a gives well-defined probabilities. However, we do not have an intuition of what is special about the measurement choices that satisfy Eqs.~\eqref{condiDef.2.a}. Moreover, the probabilities provided by definition 2a (see Table \ref{ProbDef2.a}) do not have a straightforward interpretation since they do not correspond to either ``collapse'' or full unitary evolution up to the final measurement of Wigner and unlike for definition 3 we do not find an operational principle distinguishing them. Again, a systematic study of all possible probability rules that reduce to standard quantum probabilities for non-Wigner's friend scenarios might give further insight. We leave it up for future work to do so. 

Moreover, it will be fruitful to explicitly construct maps between the perspectives of Wigner and his friend~\cite{vanrietvelde2020change} using the timeless description of the setup developed here.

In contrast to the standard quantum formalism --where there is a dichotomy between the assignment of probabilities at the moment of measurements and  unitary evolution-- the timeless formulation of quantum mechanics places the rule to assign probabilities to measurement outcomes (Born rule) logically prior to unitary dynamics. Thus, the timeless formulation is not directly concerned with the ``evolution of quantum states'' (in particular the transition from before to after a measurement), making it an appropriate tool to deal with the ambiguity of the standard formalism in describing quantum measurements.  

As long as one considers typical quantum scenarios (i.e., a system measured one or more times), there have been at least two different, yet equivalent, proposal in the timeless formulation on how to assign probabilities to the outcomes of measurements \cite{dolby2004conditional,giovannetti2015quantum}. However, we have shown that to deal with Wigner's friend scenarios these two-time probabilities rules ought to be adapted. We presented two such adaptations and considered a third proposal, all of which resolve the ambiguity of Wigner's friend probability paradox in different ways.
Thus, in this framework, the ambiguity between the two dynamical processes (i.e., unitary evolution versus ``collapse'') is pushed back to the choice of the probability-assignment rule: { Finding compelling physical arguments to rule out all but one of the proposed two-time probability-assignments would significantly contribute to solving the Wigner's friend paradox.}

\section*{Acknowledgments}
We want to thank Lorenzo Maccone for helpful discussions.
This work was funded by CoQuS, Vienna Doctoral School (VDS) and the QUOPROB project (no. I 2906- G24) of the Austrian Science Fund (FWF) and supported in part  by  the Natural  Sciences  and  Engineering  Research  Council  of Canada (NSERC) and the Dartmouth College Society of Fellows. F.D.S. acknowledges the financial support through a DOC Fellowship of the Austrian Academy of Sciences (OAW).  E.C.-R. was supported in part by the Program of Concerted Research Actions (ARC) of the Universit\'e libre de Bruxelles. 
Research at Perimeter Institute is supported in part by the Government of Canada through the Department of Innovation, Science and Industry Canada and by the Province of Ontario through the Ministry of Colleges and Universities.
C.B., further, acknowledges the support of the Austrian Science Fund (FWF) through the SFB project "BeyondC" and the project I-2526-N27, a grant from the Foundational Questions Institute (FQXi) Fund and a grant from the John Templeton Foundation (Project No. 60609). The opinions expressed in this publication are those of the authors and do not necessarily reflect the views of the John Templeton Foundation.

\bibliographystyle{unsrturl}
\bibliography{WignerCPI}

\onecolumn
\appendix

\section{Reproducing standard quantum theory in non-Wigner's friend setups }
\label{A:unitarity}
For non-Wigner's friend scenarios the two measurements are performed on the same quantum system instead of one of them measuring both the system and an observer. The constraint Hamiltonian then takes the form 
\begin{align}
\hat H' = \hat{P_t} +H_S+ \delta(\hat T - t_M) \hat{K}_{SM}+ \delta(\hat T - t_N) \hat{K}_{SN},
\label{HnonWigner}
\end{align}
where now $M$ and $N$ are apparatus or memories, in which the results of the respective measurements are encoded. This gives the physical state
\begin{align}
\ket{\Psi'}\rangle &= \int_{-\infty}^{t_M}dt \,\ket{t}U_S(t,t_0)\ket{\psi_S(t_0)}\ket {R_M}\ket{R_N} \nn \\
& \quad+  \int_{t_M}^{t_N}dt \, \ket {t}\sum_{m}  U_S(t,t_M)\Pi_m\ket{\psi_S(t_M)}\ket {m_M}\ket{R_N}  \nn\\
& \quad +  \int_{t_N}^{\infty}dt\, \ket t \sum_{m,n} \big[ U_S(t,t_N)\Pi_n  U_S(t_N,t_M)
  \Pi_m \ket{\psi_S(t_M)}\ket {m_M}\ket{n_N}\big],
\label{PsinonWigner}
\end{align}
with both $\Pi_m$ and $\Pi_n$ acting on $\mathcal{H}_S$. Moreover, we have
\begin{align} 
\bra{t}P^{\rm ph}\ket{t_0}\ket{\phi(t_0)}=\mathcal{U}(t,t_0)\ket{\phi(t_0)}
\label{PphnW}
\end{align}
for arbitrary $\ket{\phi(t_0)}\in \mathcal{H}_S\ox  \mathcal{H}_M\ox  \mathcal{H}_N$, where
\begin{align} 
&\mathcal{U}(t,t_0)=\begin{cases} 
     U_S(t,t_0) & t_0<t \\
     U_S(t,t_M)  U_{M}U_S(t_M,t_0) & t_0< t_M< t \\
     U_S(t,t_N)  U_{N}U_S(t_N,t_0) &  t_0< t_N<t \\
     U_S(t,t_N)  U_{N}U_S(t_N,t_M) U_{M}U_S(t_M,t_0)&t_0< t_M<t_N<t  \\ 
   \end{cases}
\label{unitarynW}
\end{align}
with $U_M=e^{-i \hat{K}_{SM}}$ and $U_N=e^{-i \hat{K}_{SN}}$ being the measurement unitaries that entagle the measured system with the respective memories:
\begin{align} 
U_X \ket{\psi_S}\ket{R_X}=\sum_x \Pi_x \ket{\psi_S}\ket{x_X}, \nn
\end{align}
and $(x,X)\in\{(m,M),(n,N)\}$.\\

According to definition 1 the conditional probability of result $n$ at time $t_2\geq t_N$ given result $m$ at time $t_1\geq t_M$ is
\begin{align} 
&\frac{\bbra{\Psi'} t_1\rangle \Pi^m \braket{t_1| P^{\rm ph} |t_2}\Pi^n \braket{t_2 |P\ph|t_1} \Pi^m \braket{t_1|\Psi'}\rangle}{\bbra{\Psi'}t_1\rangle \Pi^m \braket{t_1|\Psi'}\rangle},
\label{Def1nW}
\end{align}
with $\Pi^m$ and $\Pi^n$ acting on $\mathcal{H}_M$ and $\mathcal{H}_N$ respectively. From Eq.~\eqref{PsinonWigner} we  see that
$\ket{\phi (t_1)}\ce \braket{t_1|\Psi'}\rangle=
\sum_{m'}  U_S(t_1,t_M)\Pi_{m'}\ket{\psi_S(t_M)}\ket {m'_M}\ket{R_N}$,
and the denominator in Eq.\eqref{Def1nW} is
\begin{align} 
\braket{\phi(t_1)|\Pi^m|\phi(t_1)} &=\sum_{m',m''}\braket{m''_M|\Pi^m|m'_M}\bra{\psi_S(t_M)} \Pi_{m''}\Pi_{m'}\ket{\psi_S(t_M)}\nn \\
&=|\braket{m|\psi_S(t_M)}|^2.
\label{den}
\end{align}
Moreover, since
\begin{align} 
\mathcal{U}(t_2,t_1) \Pi^m\ket{\phi(t_1)} = \sum_{n'} U_S(t_2,t_N) \Pi_{b'} U_S(t_N,t_M)\Pi_{m}\ket{\psi_S(t_M)}\ket {m_M}\ket{n'_N}, 
\end{align}
the numerator in Eq.\eqref{Def1nW} gives
\begin{align} 
&\braket{\phi(t_1)|\Pi^m\mathcal{U}^\dagger(t_2,t_1) \Pi^n\mathcal{U}(t_2,t_1)\Pi^m|\phi(t_1)} \nn \\
&\qquad=\sum_{n',n''} \braket{n''_N|\Pi^n|n'_N} \bra{\psi_S(t_M)} \Pi_{m} U_S(t_M,t_N)\Pi_{n''} U_S(t_N,t_N)\Pi_{n'} U_S(t_N,t_M)\Pi_{m}\ket{\psi_S(t_M)} \nn \\
&\qquad =\braket{\psi_S(t_M)|\Pi_m U_S(t_M,t_N)\Pi_n U_S(t_N,t_M)\Pi_m|\psi_S(t_M)} \nn  \\
&\qquad =|\braket{n| U_S(t_N,t_M)|m}|^2 |\braket{m|\psi_S(t_M)}|^2
\label{num1}
\end{align}
and, hence, 
\begin{align}
P_1\left( n \ {\rm when}\ t_2 \, |\, m \ {\rm when} \ t_1 \right)=
|\braket{n|U_S(t_N,t_M)|m}|^2, 
\end{align}
which is the standard quantum probabilities for two subsequent measurements on a quantum system $S$.\\

In case of non-Wigner's friend scenarios definition 2a and definition 2b are equal since
\begin{align*} 
\bbra{\Psi'}t_2\rangle \Pi^m \braket{t_2|\Psi'}\rangle  &= \sum_{\substack{m', n\\m'',n'}}\Big[\braket{n'_N| n_N}\braket{m''_M|\Pi^m|m'_M}
\bra{\psi_S(t_M)} \Pi_{m''}U_S(t_M,t_N) \Pi_{n'}  \\[-1.5em]
&\qquad  \qquad  \cdot U_S(t_N,t_2)  U_S(t_2,t_N) \Pi_{n}  U_S(t_N,t_M) \Pi_{m'} \ket{\psi_S(t_M)}\Big]\nn \\
&= \bra{\psi_S(t_M)} \Pi_{m}U_S(t_M,t_N)\left(\sum_{n}\Pi_{n} \right) U_S(t_N,t_M) \Pi_{m} \ket{\psi_S(t_M)} \\
&=|\braket{m|\psi_S(t_M)}|^2\nn \\
&=
\bbra{\Psi'}t_1\rangle \Pi^m \braket{t_1|\Psi'}\rangle.
\end{align*}
From Eq.~\eqref{PsinonWigner} we get that
$\ket{\phi (t_2)}\ce \braket{t_2|\Psi'}\rangle =\sum_{m',n'} U_S(t,t_N)\Pi_{n'}  U_S(t_N,t_M) \Pi_{m'} \ket{\psi_S(t_M)}\ket {m'_M}\ket{n'_N}$,
and the numerator in Defs.~2 gives
\begin{align}
\langle\langle{\Psi'} \ket{t_2}\bra{t_2}\ox\Pi^{n} \ox \Pi^{m}  \kket{\Psi'}
&=\braket{\phi(t_2)| \Pi^{n} \ox \Pi^{m}|\phi(t_2)} \nn \\
&=\sum_{\substack{m', n'\\m'',n''}}\Big[\braket{n''_N|\Pi^n|n'_N}\braket{m''_M|\Pi^m|m'_M}
\bra{\psi_S(t_M)} \nn \\[-1.5em]
&\qquad  \qquad  \cdot  \Pi_{m''}U_S(t_M,t_N) \Pi_{n''}\Pi_{n'}  U_S(t_N,t_M) \Pi_{m'} \ket{\psi_S(t_M)}\Big]\nn \\
&=|\braket{b| U_S(t_N,t_M)|m}|^2 |\braket{m|\psi_S(t_M)}|^2
\label{num2}
\end{align}
Therefore, both Defs.~2 give the standard quantum probabilities
\begin{align*}
P_{2a}\left(  n \ {\rm when}\ t_2 \, |\, m \ {\rm when} \ t_1 \right)=|\braket{n|U_S(t_N,t_M)|m}|^2 =P_{2b}\left(  n \ {\rm when}\ t_2 \, |\, m \ {\rm when} \ t_2 \right).
\end{align*}

According to definition~3  the conditional probability of result $n$ at time $t_2\geq t_N$ given result $m$ at time $t_1\geq t_M$ is
\begin{align} 
&\frac{\bbra{\Psi'} t_1\rangle \Pi^m \braket{t_1| P^{\rm ph} |t_2}\Pi^n \braket{t_2|\Psi'}\rangle}{\bbra{\Psi'}t_1\rangle \Pi^m \braket{t_1|\Psi'}\rangle}.
\label{Def3nW}
\end{align}
The denominator is the same as in definition~1 and definition~2a and given by Eq.~\eqref{den}. The numerator of Eq.~\eqref{Def3nW} gives
\begin{align}
\braket{\phi(t_1)|  \Pi^m \mathcal{U}(t_1,t_2)\Pi^n|\phi(t_2)} 
&= 
\bra{R_N}\bra{m_M}\bra{\psi_S(t_M)}\Pi_{m}U_S(t_M,t_N)  U^\dagger_{N}U_S(t_N,t_2)\nn \\
&\qquad  \cdot \sum_{m'} U_S(t_2,t_N)\Pi_{n}  U_S(t_N,t_M) \Pi_{m'} \ket{\psi_S(t_M)}\ket {m'_M}\ket{n_N} \nn \\
&= \sum_{m',n'} \braket{n'_N|n_N}\braket{m_M|m'_M}
\bra{\psi_S(t_M)}\Pi_{m}U_S(t_M,t_N) \Pi_{n'}\nn \\
&\qquad \qquad \cdot \Pi_{n}  U_S(t_N,t_M) \Pi_{m'} \ket{\psi_S(t_M)}\nn \\
&=|\braket{n| U_S(t_N,t_M)|m}|^2 |\braket{m|\psi_S(t_M)}|^2.
\label{num2}
\end{align}
Hence, also definition 3 gives the standard quantum probabilities for non-Wigner's friend scenarios.

\section{Conditions for Definition 2a}
\label{A:condiDef2}

\begin{table}
\setlength{\tabcolsep}{4pt}
\renewcommand{\arraystretch}{2}
\centering
\begin{tabular}{c|c|c}
\multicolumn{3}{c}{
$\bm{\frac{\langle\langle{\Psi}\ket{t_2}\bra{t_2} \ox\Pi^{w} \otimes \Pi^{f} \kket{\Psi}}{\langle\langle{\Psi} \ket{t_1}\bra{t_1}\ox  \Pi^{f}\kket{\Psi}}}$
}  \\ 
\multicolumn{1}{c}{} &\multicolumn{1}{c}{ $\rm{yes}$ }&\multicolumn{1}{c}{$\rm{no}$ }\\ \hline 
$\ua $ & $\frac{a^2\alpha^4+2ab\alpha^3\beta \cos(\phi)+b^2\alpha^2\beta^2}{a^2}$ & $\frac{a^2\beta^4-2ab\alpha\beta^3 \cos(\phi)+b^2\alpha^2\beta^2}{a^2}$ \\
$\da$ & $\frac{b^2\beta^4+2ab\alpha\beta^3 \cos(\phi)+a^2\alpha^2\beta^2}{b^2}$ & $\frac{b^2\alpha^4-2ab\alpha^3\beta \cos(\phi)+a^2\alpha^2\beta^2}{b^2}$  \\
\end{tabular}
\caption{Definition 2a evaluated for arbitrary $a,b,\alpha,\beta$ and $\phi=\phi_S - \phi_{SF}$. These expressions constitute possible conditional probabilities of Wigner seeing result $w\in\{\rm{yes,no}\}$ at time $t_2$ given that the Friend saw result $f\in\{\ua,\da\}$ at time $t_1$ only if conditions \eqref{condiDef1} and \eqref{condiDef2} are both satisfied.}
\label{fullProb2.a}
\end{table}
Evaluating definition 2a for the setup in Fig.~\ref{circuit} gives the terms listed in Table~\ref{fullProb2.a}. Normalization requires that $\forall f:  \sum_w\frac{ \langle\langle{\Psi} \ket{t_2}\bra{t_2}\ox\Pi^{w} \otimes \Pi^{f} \kket{\Psi}\rangle }{ \langle\langle{\Psi}  \ket{t_1}\bra{t_1} \ox\Pi^{f}\kket{\Psi}}=1$, which gives the following conditions
\begin{align}
&\alpha^4+\beta^4+2\cos(\phi)(\alpha^3\beta-\alpha\beta^3)\frac{b}{a}+2\alpha^2\beta^2\left(\frac{b}{a}\right)^2=1 \label{condiDef1}\\
&\alpha^4+\beta^4-2\cos(\phi)(\alpha^3\beta-\alpha\beta^3)\frac{a}{b}+2\alpha^2\beta^2\left(\frac{a}{b}\right)^2=1.
\label{condiDef2}
\end{align}
These quadratic equations in $\frac{b}{a}$ and $\frac{a}{b}$ have solutions
\begin{align*}
&{\left(\frac{b}{a}\right)}_{\pm}=
\frac{-\cos\phi(\alpha^2-\beta^2)\pm\sqrt{1-\sin^2\phi(\alpha^2-\beta^2)^2}}{2\alpha\beta}
\end{align*}
and 
\begin{align*}
&{\left(\frac{a}{b}\right)}_{\pm}=
\frac{\cos\phi(\alpha^2-\beta^2)\pm \sqrt{1-\sin^2\phi(\alpha^2-\beta^2)^2}}{2\alpha \beta},
\end{align*}
which exist for any $\alpha,\beta$ and $\phi$. However, requiring that ${\left(\frac{b}{a}\right)}_{\pm}=1/{\left(\frac{a}{b}\right)}_{\pm}$ has solutions only when combining either the two ``$+$'' or the two ``$-$'' solutions.
These combinations are the common solutions to both  Eqs. \eqref{condiDef1} and \eqref{condiDef2} and, hence, give normalized probabilities for definition 2a. 


\section{Consistent histories for Wigner's-friend setups}
\label{A:CH}

In the consistent histories framework sequences of physical properties are assigned to a closed quantum system. These sequences are represented by tensor products of orthogonal projectors (i.e. a \emph{quantum history})
\begin{equation}
Y^{\rm i}=\rho_0\otimes P_1^{\rm{i}_1} \dots \otimes P_f^{\rm{i}_f},
\end{equation}
where $\rho_0$ is the initial state and each $P_k^{\rm{i}_k}$ corresponds to some physical property at a certain time $k$. A consistent family of histories is a complete set of histories \{$Y^{i}$\} which satisfy the consistency condition
\begin{equation}
\tr\Big(K^{\dagger}(Y^{\rm i})\rho_0 K(Y^{\rm i'})  \Big)=0 \quad \text{for } \rm{i}\neq \rm{i}'
\label{CHcondi}
\end{equation}
where $\rm{i}=(\rm{i}_1\dots \rm{i}_f)$ and $K$ is the so called chain operator defined by
\begin{equation}
K(Y^{\rm i})= P_0^{\rm{i}_1}\cdot P_0^{\rm{i}_2} \dots \cdot P_0^{\rm{i}_f},
\end{equation}
with $P_0^{\rm{i}_k}=U(t_0,t_k)P_k^{\rm{i}_k} U(t_k,t_0)$. Only within a consistent family the dynamics of quantum theory describe the respective properties over time.

For the simple Wigner's-friend setup in Fig.~\ref{circuit}, these properties are the results observed by Wigner and his Friend, $\rm{i}=(f,w)$ and $\rho_0=\ket{R_W}\ket{R_F}\proj{\psi_S}\bra{R_F}\bra{R_W}$. Condition~\eqref{condition} --under which definition 3 gives proper probabilities-- implies that
\begin{align}
\tr\Big(K^{\dagger}(Y^{(f,w)})\rho_0 K(Y^{(f',w')})  \Big) =\delta_{ff'}\delta_{ww'} \tr\big(\rho(t_2)\Pi^w U(t_2,t_1)\Pi^f U(t_1,t_2)\big),
\end{align}
where $\rho(t_2)=\mathcal{U}(t_2,t_0)\rho_0 \mathcal{U}(t_0,t_2)$. Hence, in this case, the consistency condition is satisfied. 

The solutions to the conditions on definition 2a, however, in general do \emph{not} satisfy Eq.~\eqref{CHcondi}. Consider the concrete counterexample of $\ket{\psi_S}=\sqrt{\frac{1}{2}}(\ket{\ua}+\ket{\da})$ and $\ket{\rm yes}=\alpha  \ket{\ua,\ua}+i \beta \ket{\da,\da}$. In this case one obtains
\begin{equation}
\tr\Big(K^{\dagger}(Y^{i})\rho_0 K(Y^{i'})  \Big)=\pm \delta_{ww'}\frac{i}{2}\alpha \beta \quad  \text{for } f\neq f',
\end{equation}
which still satisfies the so called weak consistency condition ${\rm Re}\left[\tr\Big(K^{\dagger}(Y^{i})\rho_0 K(Y^{i'})  \Big) \right]=0$. In contrast to the consistency condition of \eqref{CHcondi}, however, this has been shown to be highly problematic concerning trivial combination of independent subsystems as well as dynamical stability, see Ref.~\cite{diosi2004anomalies}.

\section{Conditions for Definition 3}
\label{A:condiDef3}
\begin{table}
\setlength{\tabcolsep}{8pt}
\renewcommand{\arraystretch}{2}
\centering
\begin{tabular}{c|c|c}
\multicolumn{3}{c}{
$\bm{\frac{\langle\langle{\Psi} (\ket{t_2} \bra{t_2}\otimes \Pi^{w})  P^{\rm ph} (\ket{t_1}\bra{t_1}\otimes\Pi^f ) \kket{\Psi}}{ \langle\langle\Psi \ket{t_1}\bra{t_1}\otimes \Pi^f \kket{\Psi}\rangle}}$
}  \\ 
\multicolumn{1}{c}{} &\multicolumn{1}{c}{ $\rm{yes}$ }&\multicolumn{1}{c}{$\rm{no}$ }\\ \hline 
$\ua $ & $\alpha^2 + \frac{b}{a} \alpha \beta e^{-i (\phi)}$ & $ \beta^2 - \frac{b}{a} \alpha \beta e^{-i (\phi)}$ \\
$\da$ & $\beta^2 + \frac{a}{b} \alpha \beta e^{-i (\phi)}$ & $\alpha^2 - \frac{a}{b} \alpha \beta e^{-i (\phi)}$  \\
\multicolumn{3}{c}{}\\
\multicolumn{3}{c}{
$\bm{\frac{\langle\langle{\Psi} (\ket{t_2} \bra{t_2}\otimes \Pi^{w}) P^{\rm ph}( \ket{t_1}\bra{t_1}\otimes\Pi^f)  \kket{\Psi}}{ \langle\langle{\Psi} \ket{t_2}\bra{t_2}\otimes \Pi^w \kket{\Psi}}}$
}  \\ 
\multicolumn{1}{c}{} &\multicolumn{1}{c}{ $\rm{yes}$ }&\multicolumn{1}{c}{$\rm{no}$ }\\ \hline 
$\ua $ & $ \frac{a \alpha}{a \alpha + {b \beta}e^{-i (\phi)}}$ & $\frac{b \beta}{b \beta + a \alpha e^{i (\phi)}}$ \\
$\da$ & $ \frac{a^2 \beta^2 - ab \alpha \beta e^{i (\phi)}}{ a^2 \beta^2 + b^2 \alpha^2  - 2ab \alpha \beta \cos \left( \phi\right)}$ & $ \frac{b^2 \alpha^2 - ab \alpha \beta e^{-i (\phi)}}{a^2 \beta^2 + b^2 \alpha^2  - 2ab \alpha \beta \cos \left( \phi\right)}$  \\
\end{tabular}
\caption{Definition 3 evaluated for arbitrary $a,b,\alpha,\beta$ and $\phi$ possibly constituting the conditional probabilities of Wigner seeing result $w\in\{\rm{yes,no}\}$ at time $t_2$ given that the Friend saw result $f\in\{\ua,\da\}$ at time $t_1$ and that of the Friend seeing $f$ at $t_1$ given that Wigner will see $w$ at $t_2$. These expressions are real and positive and, hence, probabilities only if conditions \eqref{1condiDef3} and either \eqref{ND1} or \eqref{ND2} are satisfied. }
\label{fullProb3}
\end{table}

Evaluating definition 3 for the setup in Fig.~\ref{circuit} gives the terms listed in Table~\ref{fullProb3}. Those expressions are in general not probabilities, since they are neither necessarily real nor positive although they do add up to one. For them to be real numbers we require that
\begin{align}
\phi=n\pi.
\label{phaseDef3}
\end{align}
For all the terms in Table~\ref{fullProb3} to be positive, of the following conditions one each must hold. If either
\begin{align}
\alpha^2-\frac{a}{b}\alpha\beta \geq0 & \ \text{ and }\ \beta^2-\frac{b}{a}\alpha\beta \geq0\\
\label{1condiDef3}
&\ \text{ or }\ \nn \\
\alpha^2-\frac{b}{a}\alpha\beta \geq0 &\ \text{ and }\  \beta^2-\frac{a}{b}\alpha\beta \geq0,
\end{align}
 the conditional probabilities for result $w$ at $t_2$ given result $f$ at $t_1$ are well-defined. And if, either
\begin{align}
1-\frac{b\alpha}{a\beta} \geq0 &\ \text{ and }\ 1-\frac{a\beta}{b\alpha} \geq0\\
&\ \text{ or }\ \nn \\
1-\frac{b\beta}{a\alpha} \geq0 &\ \text{ and }\ 1-\frac{a\alpha}{b\beta} \geq0, 
\label{2condiDef3}
\end{align}
the conditional probabilities for result $f$ at $t_1$ given result $w$ at $t_2$ are well-defined.
The solutions to both cases are
\begin{align}
a=\alpha\quad&,\quad b=\beta  \label{ND1} \\ 
&\text{or} \nn \\
a=\beta  \quad &, \quad b=-\alpha, \label{ND2}
\end{align}
which means Wigner's measurement is aligned with the initial state. The joint system of $F$ and $S$ is in an eigenstate of $W$'s observable, i.e., the measurement is non disturbing.

\end{document}